\documentclass[aps, prb, notitlepage, twocolumn, superscriptaddress, 10pt, longbibliography]{revtex4-2}
\usepackage{color}
\usepackage{amsmath, amsthm, amssymb, bbold}
\usepackage{lipsum}

\usepackage{graphicx}
\usepackage{framed}
\usepackage[normalem]{ulem}
\usepackage{cancel}
\usepackage[version=4]{mhchem}

\usepackage{bm}

\usepackage{microtype}

\usepackage{hyperref}
\hypersetup{
	colorlinks=true,
	linkcolor=blue,
	citecolor=blue,
	urlcolor=blue
}

\newcommand{\hc}{{\rm H.c.}}
\newcommand{\up}{\uparrow}
\newcommand{\down}{\downarrow}
\newcommand{\avg}[1]{\langle #1\rangle}
\newcommand{\mc}[1]{\mathcal{#1}}

\def\comment#1{}
\def\a{\alpha}
\def\b{\beta}
\def\g{\gamma}
\def\e{\varepsilon}
\def\re{\text{Re}}
\def\im{\text{Im}}
\def\tr{\text{Tr}}
\def\sgn{\mathrm{sgn}}

\begin{document}
\title{Semiclassical analysis of spin dynamics in the non-Hermitian Hubbard model}
\author{Doru Sticlet}
\email{doru.sticlet@itim-cj.ro}
\affiliation{National Institute for R\&D of Isotopic and Molecular Technologies, 67-103 Donat, 400293 Cluj-Napoca, Romania}
\author{C\u{a}t\u{a}lin Pa\c{s}cu Moca}
\email{mocap@uoradea.ro}
\affiliation{Department of Theoretical Physics, Institute of Physics, Budapest University of Technology and Economics, 	M\H{u}egyetem rkp.~3, H-1111 Budapest, Hungary}
\affiliation{Department  of  Physics,  University  of  Oradea,  410087,  Oradea,  Romania}
\author{Bal\'azs D\'ora}
\email{dora.balazs@ttk.bme.hu}
\affiliation{Department of Theoretical Physics, Institute of Physics, Budapest University of Technology and Economics, M\H{u}egyetem rkp.~3, H-1111 Budapest, Hungary}

\begin{abstract}
We investigate a specific limit of the one-dimensional non-Hermitian Hubbard Hamiltonian with complex interactions. 
In this framework, fermions with different spin quantum numbers are mapped onto two distinct spin species, resulting in two $XY$ spin chains that are coupled through Ising $ZZ$ interaction. 
The spin ladder model is then examined within the semiclassical limit using a spin-coherent state basis, where the dynamics is governed by a set of coupled Landau-Lifshitz-Gilbert equations.
The non-Hermitian interactions in this model generate a spin-transfer torque term.
We analyze the system's evolution toward several potential steady states, including a state of decoupled chains that is accessible when each chain has uniform initial conditions. 
Other possible steady states involve dimerized configurations with decoupled rungs, where the rung spins are either ferromagnetically or antiferromagnetically coupled, depending on the sign of the imaginary interactions. 
The spin dynamics is then studied in the infinite-temperature limit, which favors dimerized steady states.
Despite the decoupled rungs, we observe the formation of ferromagnetic domains along each chain in the steady state. 
Additionally, we investigate the spin correlation functions and identify signatures of anomalous spin dynamics.
\end{abstract}

\maketitle
\section{Introduction}

The field of non-Hermitian physics has experienced significant growth in recent years, driven by both theoretical predictions and experimental discoveries across various platforms, including optics, acoustics, and ultracold atomic gases~\cite{Bender2007,Moiseyev2011,ElGanainy2018,Ashida2020,Bergholtz2021,Okuma2023}. 
This surge of interest has unveiled a range of intriguing phenomena, such as spectral exceptional points~\cite{Berry2004,Heiss2012,Miri2019}, chiral transport around these points~\cite{Uzdin2011,Berry2011,Doppler2016}, the non-Hermitian skin effect~\cite{Yao2018,Yao2018a}, and $PT$-symmetric systems with real eigenvalues~\cite{Bender1998,Makris2008}, among others. 
These phenomena are typically understood within the framework of single-particle physics, which has led to a topological classification of non-interacting non-Hermitian Hamiltonians~\cite{Shen2018,Lieu2018,Kawabata2019}. 
As we advance, the focus of research is shifting towards many-body systems, where the interplay between interactions and non-Hermiticity promises to uncover even more complex and surprising phenomena.

In this context, the study of the Hubbard model, a fundamental example of strongly correlated systems~\cite{Essler2005,Arovas2022}, naturally emerges as a key area of exploration. 
The Hubbard model has been extended into the non-Hermitian realm in two primary (and not mutually exclusive) ways. 
The first approach involves starting with the Hatano-Nelson model~\cite{Hatano1996}, which features non-reciprocal hopping terms, and incorporating traditional (Hermitian) on-site Hubbard interactions~\cite{Fukui1998}. 
In these models, various phenomena have been explored, including the skin effect~\cite{Zheng2024}, full-counting statistics~\cite{Dora2022}, transitions at exceptional points~\cite{Zhang2022}, many-body localization~\cite{Hamazaki2019,Heussen2021}, topological properties~\cite{Kawabata2022}, quantum quenches~\cite{Dora2023}, and entanglement dynamics~\cite{Orito2023}.

The second approach involves extending Hubbard on-site interactions to complex values, which is the focus of the present work.
Such models arise in the context of effective non-Hermitian Hamiltonians within the Lindblad formalism for open quantum systems, and in the theoretical analysis of hopping models with dephasing noise~\cite{Medvedyeva2016}. 
They also occur in physical scenarios involving two-particle losses in both fermionic and bosonic systems~\cite{Syassen2008,FossFeig2012,Zhu2014,Ashida2016,Tomita2017,Nakagawa2018,Sponselee2018,Yamamoto2019,Nakagawa2020,Yamamoto2021,Yamamoto2022,Rosso2022,Rosso2023,Mazza2023}. 
In specific cases, where the model features pure gain or loss, it may be exactly solvable, providing insights into the Liouvillian eigenspectrum~\cite{Buca2020,Nakagawa2021,Yoshida2023}.

In this work, we examine the semiclassical limit of the non-Hermitian Hubbard model with imaginary interactions, which is derived by mapping the system onto a spin ladder model and considering the limit of large spins. 
Our objective is to explore the spin dynamics within this semiclassical limit, determine if the system evolves to a steady state (SS), and characterize the properties of such a state.

This investigation is timely given the recent resurgence of interest in the dynamics of classical spin models. 
The classical Heisenberg chain model has demonstrated diffusive spin dynamics in the infinite-temperature limit~\cite{Gerling1990,Bagchi2013}. 
Furthermore, classical spin models have recently been studied for their anomalous dynamics, including signatures of Kardar-Parisi-Zhang scaling~\cite{Kardar1986}, at both finite and infinite temperatures~\cite{DeNardis2018,Das2018,Das2019,Li2019,Bulchandani2021,McRoberts2022,McRoberts2022a,Roy2023,Krajnik2024}.

The study of non-Hermiticity in classical spin chains has received less attention. However, it has been demonstrated that incorporating non-Hermiticity into classical spin models introduces damping effects in the Landau-Lifshitz equations through spin-torque terms~\cite{Wieser2013,Wieser2015,Galda2016}. 
Subsequent work has explored the inverse approach, deriving non-Hermitian effective Hamiltonians for magnons from Landau-Lifshitz-Gilbert equations, with a focus on topology and exceptional points in classical spin systems~\cite{Tserkovnyak2020,Flebus2020,Hurst2022,Zou2024,Yu2024}.

The classical spin model examined here, derived from the Hubbard model, consists of two $XY$ spin chains coupled via Ising $ZZ$ interactions.
We use local coherent states to solve the non-Hermitian model, providing a mean-field description similar to the analysis of a non-Hermitian Bose-Hubbard dimer~\cite{Graefe2008}. 
It has been established previously~\cite{Bari1973} that the Hermitian version of this model shows a metal-insulator transition depending on the ratio of interaction to hopping strengths.
In contrast, our results reveal that the steady state of the infinite-temperature non-Hermitian model is an Ising-like state with spins polarized along the $z$ axis. 
In this steady state, the rungs of the spin ladder exhibit either ferromagnetic or antiferromagnetic coupling, contingent upon the sign of the imaginary interactions.  Moreover, initial correlations along the chain during the transient regime can lead to the formation of domain walls.
While the typical steady state exhibits dimerized rungs, alternative steady states can arise under specific conditions for uniform initial states. 
In these cases, the two chains decouple, leading to a steady state characterized by a real energy spectrum.

The organization of the article is the following. 
In Sec.~\ref{sec:quantum}, we introduce the non-Hermitian quantum Hubbard model, map it to a spin model, and obtain the equations of motion for the spin expectation values.
In Sec.~\ref{sec:semiclassical}, we take the semiclassical limit of large spins, and derive from the quantum model the Landau-Lifshitz-Gilbert equations governing the spin dynamics. 
Section~\ref{sec:uniform} analyzes the equations of motion with uniform initial conditions where the phase diagram is analytically determined. 
Sec.~\ref{sec:infinite} examines the infinite-temperature limit of the classical model. 
Here, we examine how non-Hermitian interactions determine the steady state and the characteristic time required to reach it. 
We demonstrate that the preferred steady state is a dimerized configuration with either antiferromagnetic or ferromagnetic coupling between rung spins.
In Sec.~\ref{sec:dynamic}, we uncover signatures of anomalous spin dynamics in the spin-spin correlation functions.
Finally, we discuss the results in Sec.~\ref{sec:conclusion}.

\section{Models}
\subsection{Quantum model and equations of motion}
\label{sec:quantum}
The starting point of our analysis is the non-Hermitian version of the one-dimensional Fermi-Hubbard model 
\begin{align}\label{FH}
H=&-{\cal J} \sum_{j\sigma} (c^\dag_{j\sigma} c^{}_{j+1\sigma}+\hc)
-\text{\textmu}\sum_j(n_{j\up}+n_{j\down})\notag\\
&+{\cal U}e^{i\phi}\sum_j (n_{j\up}-\frac12) (n_{j\down}-\frac12).
\end{align}
The fermion creation and annihilation operators are denoted by $c_{j\sigma}$ and $c_{j\sigma}^\dag$, respectively, with site $j$ and spin $\sigma$ indices.
The hopping amplitude is parameterized by ${\cal J}$, and {\textmu} is the chemical potential.
The on-site Coulomb interaction is complex, and it is parameterized by an amplitude ${\cal U}$ and a phase $\phi$. 
Without loss of generality, we take ${\cal U}>0$ and $\phi\in[0,2\pi)$. 
The Hermitian Hubbard model with repulsive or attractive interactions is realized for $\phi=0$ and $\phi=\pi$, respectively.
The non-Hermitian Hamiltonian \eqref{FH} naturally arises as an effective Hamiltonian describing the short-time dynamics in Lindbladian evolution when two-body losses are taken into account, and the recycling term is ignored~\cite{Nakagawa2020, Ashida2020}. In this context, the imaginary part of the interaction $\mc{U}\sin\phi$ is proportional to the two-body loss rate~(see App.~\ref{sec:lindblad}).

The model is mapped to a spin Hamiltonian with two spin species $S_j^\alpha$ and $T^\alpha_j$ (lower index for site $j$ and upper index for spin projection) using the Jordan-Wigner transformation,
\begin{gather}
c_{j\up}=e^{i\pi\sum_{k=1}^{j-1}n_{k\up}}S_j^-,\quad 
c_{j\down}=e^{i\pi\sum_{k=1}^N n_{k\up}}
e^{i\pi\sum_{k=1}^{j-1}n_{k\down}}T_j^-,
\end{gather}
with $S^\pm_j=S^x_j\pm S^y_j$ and $T^\pm_j=T^x_j\pm T^y_j$. 
The resulting spin Hamiltonian reads~\cite{Bari1973,Shastry1986}
\begin{align}\label{FH_spin}
H=& -{\cal J}\sum_j (S^+_jS^-_{j+1} + T^+_jT^-_{j+1} + \hc) + {\cal U}e^{i\phi}\sum_jS_j^zT_j^z \notag\\
&-\text{\textmu}\sum_j(S^z_j+T^z_j+1),
\end{align}
with $S^z_j=n_{j\up}-1/2$, $T^z_j=n_{j\down}-1/2$. 
The Hubbard model is thus mapped onto a ladder with $XY$ couplings for the spins along the rails and Ising-like $ZZ$ couplings on the rungs of the ladder.


The dynamics of the expectation values for the spin components are obtained using the non-Hermitian Heisenberg equation~\cite{Graefe2008, Sticlet2022}
\begin{equation}\label{eom1}
i\hbar\frac{d}{dt}\avg{S^\a_j} = 
\avg{S^\a_j H-H^\dag S^\a_j} - \avg{S^\a_j}\avg{H-H^\dag},
\end{equation}
with 
\begin{equation}\label{expect_val}
\avg{S_j^\alpha} = \tr[\rho(t)S^\alpha_j]/\tr[\rho(t)],
\end{equation}
and $\rho(t)$ the time-dependent density matrix.
A similar equation holds for $\avg{T_j^\alpha}$ evolution.
This formulation is standard for describing dissipative dynamics in the absence of quantum jumps~\cite{Dalibard1992, Daley2014}. In such cases, the density matrix is typically expressed in a basis of the system's right eigenvectors, and should not be confused with the biorthogonal density matrix~\cite{Herviou2019}.
It is also useful to rewrite Eq.~\eqref{eom1} by explicitly separating the Hamiltonian into its Hermitian and anti-Hermitian parts $H=H_+ + i\varGamma$,
\begin{equation}\label{eom2}
i\hbar\frac{d}{dt}\avg{S^\a_j} = 
\avg{[S^\a_j,H_+]} +i\avg{\{S^\a_j,\varGamma\}} - 2i\avg{S^\a_j}\avg{\varGamma},
\end{equation}
with $[.,.]$ and $\{.,.\}$ denoting the commutator and, respectively, the anticommutator.
Several difficulties render the quantum problem challenging.
In contrast to the Hermitian case, Eq.~\eqref{eom2} reveals that the non-Hermitian terms, through the anticommutator, generate three-spin correlation functions in the dynamics. 
Additionally, the expectation values tend to diverge due to the non-unitary evolution of the system's density matrix. Thus, similar to the quantum trajectory method used in open quantum systems~\cite{Daley2014}, explicitly normalizing the time-dependent density matrix in Eq.~\eqref{expect_val} is crucial. However, addressing these challenges is beyond the scope of this work, as we will focus on the semiclassical limit in what follows.

\subsection{Semiclassical limit}
\label{sec:semiclassical}
In the semiclassical limit, the two spin species have the same amplitude $S$, which is considered large, $S\to \infty$. 
At the same time, $\hbar \to 0$, so that $\hbar S$ remains finite.
The expectation values in Eq.~\eqref{eom2} are now evaluated by assuming that the system wave function is a direct product of coherent states of individual spins~\cite{Radcliffe1971,Schliemann1998} (see App.~\ref{sec:coherent} for details).
Within this variational ansatz, they factorize into single-spin expectation values and yield corrections to order $S$ to the classical limit.
Since the two spin species are at different lattice positions, the spins commute, and  expectation values factorize trivially within the variational ansatz,
\begin{equation}\label{fact1}
\avg{S^\a_i T^\b_j} = \avg{S^\a_i}\avg{T^\b_j}.
\end{equation}
When evaluating Eq.\eqref{eom2}, only two-spin expectation values remain. 
Similarly, $\avg{S^\a_i S^\b_j}$ and $\avg{T^\a_i T^\b_j}$ factorize as in Eq.~\eqref{fact1} when the sites $i$ and $j$ are different.
However, the non-commutativity of quantum spins at the same site is reflected in the spin-coherent states.
The averages of the form  $\avg{S^\a_j S^\b_j}$ are evaluated as 
\begin{align}\label{fact2}
\avg{S^\a_j S^\b_j} = \avg{S^\a_j}\avg{S^\b_j}\Big(1-\frac{1}{2S}\Big)+\delta_{\a\b}\frac{S}{2}
+i\e_{\a\b\g}\frac{\avg{S^\g_j}}{2},
\end{align}
where $\e_{\a\b\g}$ is the Levi-Civita symbol and $\delta_{\a\b}$ is the Kronecker delta, and index $\gamma$ is summed over.
The expectation values of $\avg{T^\a_j T^\b_j}$ factorize identically. 
Within the classical limit, one would be tempted to disregard the corrections of order $S$, corresponding to the second and third terms in Eq.~\eqref{fact2}, as they appear as subdominant contributions with respect to the leading $\sim S^2$ term.
Nevertheless, the commutator $\avg{[S^\alpha_j, H_+]}$ in Eq.~\eqref{eom2} singles out the linear in $S$ contribution, $i\e_{\a\b\g}\avg{S^\g_j}$, yielding the usual Poisson structure in the equations of motion.
Moreover, within the anticommutator with the non-Hermitian part, the term $\sim S^2$ from Eq.~\eqref{fact2}, cancels against the $-2i\avg{S^\a_j}\avg{\varGamma}$ contribution.
Thus, both commutator and anticommutator terms in Eq.~\eqref{eom2} give contributions of the order $S^2$ [since also the term $\avg{T^z_j}$ contributes to order $S$], so the expression~\eqref{fact2} needs to be used without discarding subleading terms. 

In the semiclassical limit, rescaling of the classical spins is necessary to ensure their amplitudes are properly normalized. 
Let us introduce the classical spins of unit norm,
\begin{equation}\label{unitspins}
    s_j^\a = \frac{\langle S_j^\a \rangle}{S},\quad \tau_j^\a = \frac{\langle T_j^\a \rangle}{S}.
\end{equation}
Moreover, the diverging spin amplitudes are absorbed into the exchange coupling ${\cal J}$, the Coulomb energy $\cal U$
and the chemical potential $\text{\textmu}$
\begin{equation}
    \bar{J} = \mc{J}S^2, \quad \bar{U} = {\cal U} S^2, \quad {\bar \mu} = \text{\textmu} S,
\end{equation}
which remain finite as $\hbar \to 0$. 
Additionally, it is more appropriate to express energy in frequency units by rescaling the energy with $\hbar S$. 
This leads us to introduce the parameters of the classical model,
\begin{gather}
    J = \frac{{\bar J}}{\hbar S},\quad U = \frac{{\bar U}}{\hbar S},\quad \mu = \frac{{\bar \mu}}{\hbar S},
\end{gather}
in terms of which the rescaled Hamiltonian, $\mathcal H  = \avg{H}/\hbar S$, is expressed as
\begin{align}\label{classical_H}
\mathcal H = & -2J\sum_j (s_j^xs_{j+1}^x +s_j^y s_{j+1}^y+\tau_j^x\tau_{j+1}^x +\tau_j^y \tau_{j+1}^y)\notag\\ 
&+ U e^{i\phi}\sum_j s_j^z \tau_j^z-\mu\sum_j(s^z_j+\tau^z_j+\frac1S).
\end{align}
In the following, the last term in Eq.~\eqref{classical_H} will be dropped out, since it is irrelevant as $S\to\infty$.
Using Eqs.~\eqref{fact1} and \eqref{fact2} into the quantum equations of motion~\eqref{eom2}, evaluated in a spin-coherent basis, yields after some algebra~[App.~\ref{sec:eoms}] the semiclassical equations of motion for the rescaled spins 
\begin{align}\label{classical_eoms}
\dot{\bm s}_j &= \frac{d\re\mc H}{d\bm s_j}\times \bm s_j
- \bigg(\frac{d\im\mc H}{d\bm s_j}\times\bm s_j\bigg)\times \bm s_j,\notag\\
\dot{\bm \tau}_j &= \frac{d\re\mc H}{d\bm \tau_j}\times \bm \tau_j 
- \bigg(\frac{d\im\mc H}{d\bm \tau_j}\times\bm \tau_j\bigg)\times \bm \tau_j,
\end{align}
with $\re\mc H = \avg{H_+}/\hbar S$ and $\im\mc H = \avg{\varGamma}/\hbar S$.
The commutator with the Hermitian part $H_+$ of the Hamiltonian ~\eqref{classical_H} yields the familiar Landau-Lifshitz equations of motion, corresponding to the first term on the right-hand side of Eqs.~\eqref{classical_eoms}. 
Additionally, the anticommutator with the anti-Hermitian term $\varGamma$ generates a Slonczewski spin-transfer torque, represented by the second term on the right-hand side of Eqs.~\eqref{classical_eoms}. 
The appearance of spin-transfer torque due to a non-Hermitian Hamiltonian aligns with previous studies that examined the non-Hermitian dynamics of single spins~\cite{Wieser2013,Galda2016}.

The Eqs.~\eqref{classical_eoms} possess several advantageous properties. Despite the classical energy~\eqref{classical_H} being imaginary, the equations of motion are entirely real. 
Additionally, these equations are symmetric with respect to the interchange of spins $\bm s$ and $\bm \tau$. 
Finally, the norm of the classical spins is conserved throughout the evolution, as evident from Eqs.~\eqref{classical_eoms}.

To analyze these equations, the $z$ component plays a crucial role in determining the steady state. It is helpful to introduce the magnetization $m_j$ at a rung $j$, the total magnetization $m$, and the corresponding staggered magnetization $l_j$, along with the total staggered magnetization $l$,
\begin{align}\label{magnetization}
m_j &= \frac{1}{2}(s^z_j+\tau^z_j), & m &= \frac{1}{N}\sum_j m_j,\notag\\
l_j &= \frac{1}{2}(s^z_j-\tau^z_j), & l &= \frac{1}{N}\sum_j l_j,
\end{align}
as quantities that describe the system's dynamics.

\section{Uniform initial conditions}
\label{sec:uniform}
We begin by analyzing the case in which the spins in each chain start in a uniform state. This simplifies the problem to a minimal two-spin scenario (or a single-site Hubbard model in the original quantum context), where a periodic spin ladder with all spins $\bm{s}_j$ initially in the same arbitrary state $\bm{s}(0)$ and all $\bm{\tau}_j$ spins in state $\bm{\tau}(0)$ is considered. 
Due to spatial translation invariance, spins on different rungs evolve identically, allowing us to disregard the site index.
This reduces Eqs.~\eqref{classical_eoms} to six coupled equations for the two spins and their three components, $\bm{s}(t)$ and $\bm{\tau}(t)$. 
Since the norm of the spins is conserved during their evolution, there are only four independent variables, which can be represented by the angles determining the spins' orientations on the two Bloch spheres. 
However, we find that the $z$ components play a special role, making it more convenient to retain the full formulation in terms of the three spin components.

\begin{figure}[t]
    \includegraphics[width=\columnwidth]{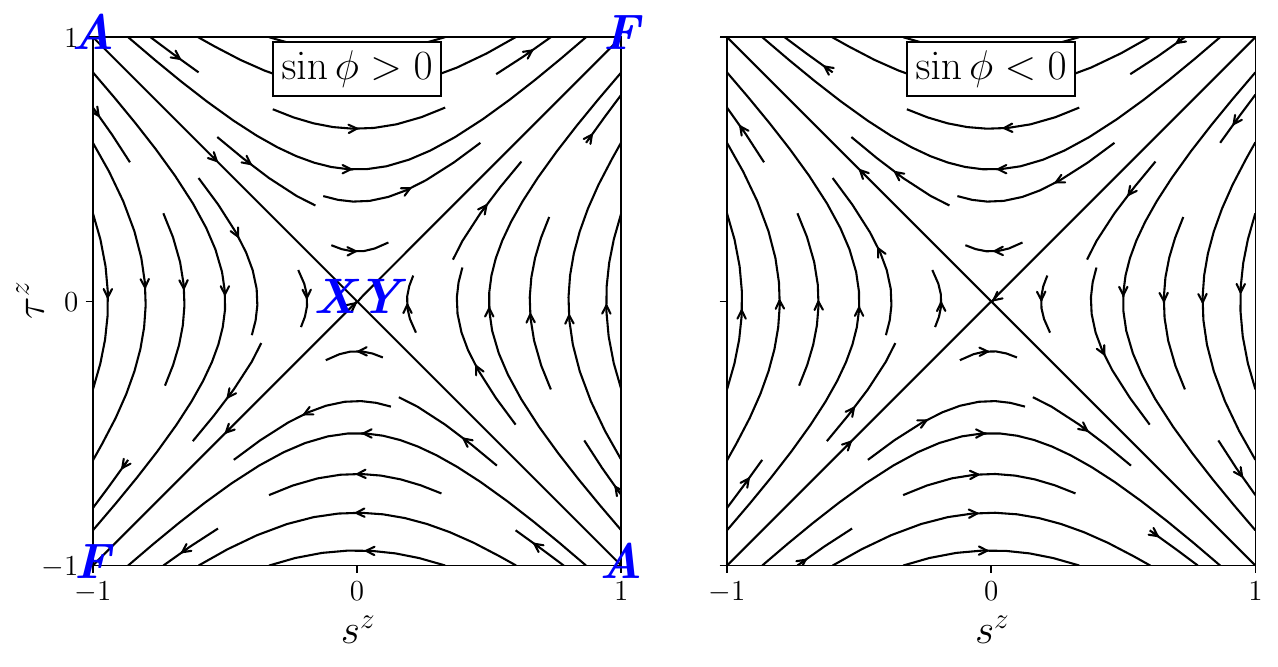}
    \includegraphics[width=\columnwidth]{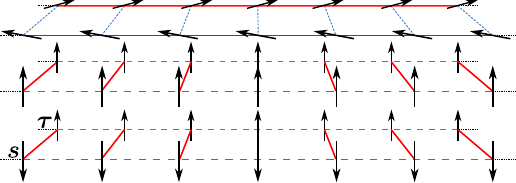}
    \caption{(Top) Flow diagram showing the evolution towards the steady states for the uniform chain with (left) positive $\sin\phi > 0$ and (right) negative imaginary interactions $\sin\phi < 0$. 
    The steady states are indicated on the left figure with letters. The dimerized SSs are denoted by $A$, for antiferromagnetic, or $F$, for ferromagnetic rungs, while $XY$ indicates the $XY$ SS, corresponding to the Hermitian point.
    (Bottom) The three qualitatively different steady states in the uniform system. 
    The $XY$ SS has coupling (in red) only along the chains with some arbitrary orientation of spins in the equatorial plane, but no inter-chain coupling (dashed blue line). 
    The ferromagnetic or antiferromagnetic SSs have spins polarized along the $z$, coupling between rungs, and no intra-chain coupling.}
    \label{fig:uniformPhase}
\end{figure}

In the uniform case, the equations for the $z$ components are decoupled from the others, and they are sufficient to analytically determine the steady states of the system. 
From Eqs.~\eqref{classical_eoms}, we find
\begin{equation}
\dot s^z = -  U\sin\phi[(s^z)^2-1]\tau^z,
\end{equation}
and an equivalent equation for $\tau^z$ by interchanging $s^z$ and $\tau^z$. To reach the steady state
it is necessary that $\dot{s}^z = 0$ and $\dot{\tau}^z = 0$, which results in
\begin{equation}
[(s^z)^2-1]\tau^z =0,\quad [(\tau^z)^2-1]s^z=0.
\end{equation}
There are two classes of steady states. In the first, $s^z = 0$ and $\tau^z = 0$, the spins lie in the  plane, and the two chains are completely decoupled. This steady state, referred to as the $XY$ steady state, corresponds to two non-interacting fermionic chains in the original quantum model. Since the non-Hermitian interactions vanish, this is a Hermitian phase.
The second steady state occurs when the spins are fully polarized along the $z$-axis, with $|s^z| = |\tau^z| = 1$. Due to norm conservation, the $x$ and $y$ components of the spins are zero, causing the ladder rungs to be entirely decoupled from one another. We refer to this state as the dimerized SS or $ZZ$ SS. 
It has two variations: the ferromagnetic arrangement, where the spins along the rungs are ferromagnetically aligned, $s^z = \tau^z$, and the antiferromagnetic arrangement, where $s^z = -\tau^z$. 
The preferred steady state is determined by the sign of the imaginary interactions. Typical spin arrangements in the two steady states are illustrated in Fig.~\ref{fig:uniformPhase} (bottom).
The spin dynamics near the steady states is analyzed by linearizing the equations of motion.
Close to the $XY$ SS, a small deviation in the $z$ spin projections, $\delta s^z$ and $\delta \tau^z$, gives
\begin{equation}
\delta \dot s^z = U\sin\phi\delta\tau^z,\quad \delta\dot\tau^z = U\sin\phi\delta s^z.
\end{equation}
The equations become decoupled when expressed in terms of magnetization and staggered magnetization, as in Eqs.~\eqref{magnetization}, with $m(0)$ and $l(0)$ near the $XY$ point. The deviation from the $XY$ SS provides the solutions
\begin{equation}
\label{equator_ml}
m(t)=m(0)e^{U\sin(\phi)t},\quad
l(t)=l(0)e^{-U\sin(\phi)t}.
\end{equation}
The $XY$ SS is generally unstable except under specific initial conditions. 
In particular, the $XY$ point is attractive when starting with initial antiferromagnetic conditions where $m(0)=0$ and $l(0)\neq 0$. 
In this case, the magnetization remains zero at any finite time, and since $l(t_\infty) \to 0$, the system eventually flows to $s^z = \tau^z = 0$. 
Here and in the following, we denote $t_\infty$ as the time at which the system reaches its steady state.

For other initial conditions, with positive imaginary interactions $\sin\phi > 0$, the staggered magnetization vanishes, indicating that the system flows to the ferromagnetic point with spins polarized along the $z$ axis and coupled ferromagnetically ($s^z = \tau^z$). Conversely, with negative imaginary interactions $\sin\phi < 0$, the roles of $m$ and $l$ are reversed: the magnetization $m$ vanishes, suggesting that the spins tend to align in an antiferromagnetic configuration ($s^z = -\tau^z$). The only exception is when the initial conditions are $l(0) = 0$ and $m(0) \neq 0$, in which case the system flows to the $XY$ point.

Near the antiferromagnetic and ferromagnetic points, the equations for the $z$ components are linearized to confirm the stability analysis near the $XY$ point. 
For example, near the ferromagnetic point ($s^z = \tau^z = \pm 1$), one considers $s^z(t) = \pm 1 + \delta s^z(t)$ and $\tau^z(t) = \pm 1 + \delta \tau^z(t)$, leading to the equations:
\begin{equation}\label{ferropoint}
\delta \dot{s}^z = -2U \sin\phi \, \delta s^z, \quad \delta \dot{\tau}^z = -2U \sin\phi \, \delta \tau^z.
\end{equation}
The solution shows that the ferromagnetic point is attractive for positive imaginary interactions ($\sin\phi > 0$) and repulsive for negative imaginary interactions ($\sin\phi < 0$). 
Similarly, the antiferromagnetic point is attractive for negative imaginary interactions ($\sin\phi < 0$) and repulsive for positive imaginary interactions ($\sin\phi > 0$).

Numerical solutions of the equation of motion for the $z$ component reveal that the flow evolves smoothly between the analytical solutions near the $XY$ point and the ferromagnetic ($F$) and antiferromagnetic ($A$) points. 
The flow diagram, shown in Fig.~\ref{fig:uniformPhase}(top), illustrates both attractive and repulsive interactions. This confirms the general tendency of the system to evolve to the dimerized SSs, with ferromagnetic rungs at positive $\sin\phi$, and antiferromagnetic ones at negative $\sin\phi$. 

\begin{figure}[t]
\includegraphics[width=\columnwidth]{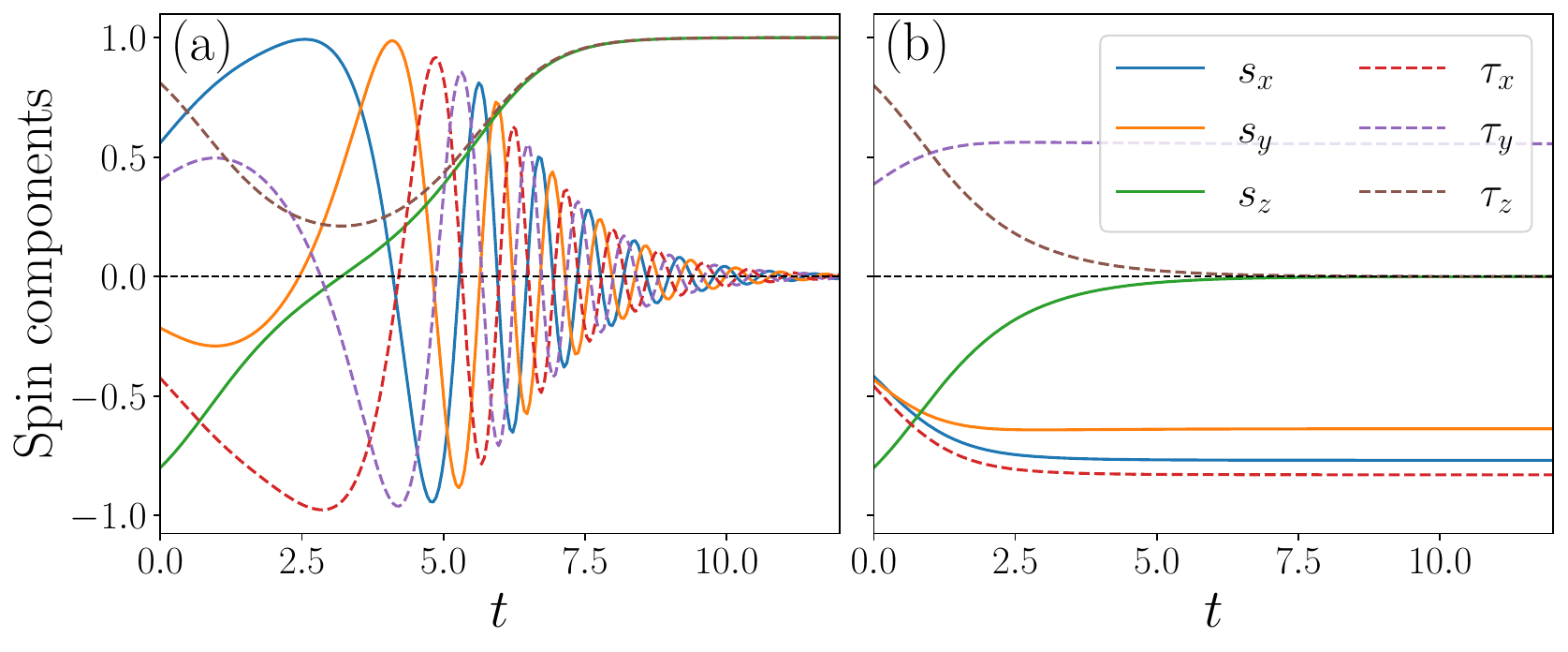}
\caption{Typical evolution of spin components (solid line, $\bm s$, and dashed line, $\bm \tau$ components) in the uniform system at positive imaginary interactions, $U=4$, $\phi=0.2$, and $\mu=0$. (a) For generic conditions, here $s^z(0)=0.8$, $\tau^z(0)=-0.81$, and random $xy$ components, the system flows to the ferromagnetic point with $s^z(t_\infty)=\tau^z(t_\infty)=1$. (b) The $XY$ point is attractive at $\sin\phi>0$ for antiferromagnetic initial conditions, here $s^z(0)=-\tau^z(0)=0.8$, and random $xy$ components. 
From the initial antiferromagnetic rung ordering, the $z$ components show exponential decay.
}
\label{fig:uniform_evol}
\end{figure}

The $XY$ point is attractive only for specific parameter choices, such as when spins are initially arranged antiferromagnetically at $\sin\phi > 0$ or ferromagnetically at $\sin\phi < 0$. 
In most cases, apart from these special scenarios, the system evolves to a state where the imaginary part of the energy is maximized, given by $\mc H = iNU |\sin\phi|$ from Eq.~\eqref{classical_H}. 
For instance, at $\sin\phi > 0$, the steady state may either be the ferromagnetic one, as shown in Fig.~\ref{fig:uniform_evol}(a), or the $XY$ SS, depicted in Fig.~\ref{fig:uniform_evol}(b), depending on the initial conditions. Unlike the $F$ or $A$ steady states, the $XY$ SS has a real energy spectrum $\mc H = -4NJ$, according to Eq.~\eqref{classical_H}.

A small periodic perturbation around the steady state is analyzed within the linear spin wave theory. 
We examine normal mode solutions near the ferromagnetic and antiferromagnetic points to derive the magnon spectrum. 
In this approach, $s^z$ and $\tau^z$ are approximated by their steady-state values, $s^z \simeq \pm 1$ and $\tau^z \simeq \pm 1$. 
Consequently, the dispersion relation for magnons in the $\bm{s}$ chain is directly obtained from the dynamics of $s^x$ and $s^y$ as described by Eqs.~\eqref{classical_eoms},
\begin{align}
\omega_\pm=&-iU\sin(\phi)\sgn(s^z\tau^z)\notag\\
&\pm|4J\cos(q)\sgn(s^z)+U\cos(\phi)\sgn(\tau^z)-\mu|,
\end{align}
and similarly for the $\bm\tau$ chain under the interchange $s^z\leftrightarrow \tau^z$.
The chemical potential induces an imbalance between the two chains, and the two magnon frequencies are equal at $\mu=0$.

As expected, for positive $\sin\phi > 0$, the spin waves are damped near the ferromagnetic point, with the amplitude of the $x$ and $y$ components decaying to zero as $\sim \exp[-U \sin(\phi) t]$. For negative $\sin\phi < 0$, the magnon frequency is similarly damped near the antiferromagnetic point, as the $x$ and $y$ components decay according to the same exponential law. 
This result concurs with the linearized equations of motion near the stationary points.
In contrast, when approaching the $F$ or $A$ steady states, the $x$ and $y$ components of the spin diverge, indicating instability at the anti- and ferromagnetic points. Trajectories on the Bloch sphere move towards the poles depending on the sign of the imaginary interactions. 

However, magnons are damped as expected when approaching the $F$ or $A$ steady states. The only exceptions occur when trajectories remain in the equatorial plane, associated with the $XY$ SS. For small deviations in the $x$ and $y$ components and appropriate initial conditions (i.e., $\sin\phi > 0$ with $m(0) = 0$ and $l(0) \neq 0$, or $\sin\phi < 0$ with $m(0) \neq 0$ and $l(0) = 0$), the trajectories stay on the equator of the Bloch sphere, leading to exponential decay in the $z$ components. 
Once on the equator, the spins rotate in the $xy$ plane with a frequency $\omega_\pm = \pm |\mu|$ [see Eq.~\eqref{hom_6}]. In that regard, for finite $\mu$, the $XY$ SS is not a true steady state, as the spins continuously rotate in the equatorial plane. 

Although this section focuses on the uniform case, the analysis also applies to random initial conditions for the $x$ and $y$ components, provided the $z$ components are uniform.

\section{Infinite temperature limit}
\label{sec:infinite}

Let us now consider the infinite-temperature limit, $ T \to \infty $, where spin translation symmetry is broken due to the random initial spins at each site. 
The nature of spin dynamics in this limit was the subject of considerable debate in the classical Heisenberg chain spin model until diffusive dynamics was finally confirmed numerically~\cite{Mueller1988,Gerling1990,AlcantaraBonfim1992,Boehm1993,Bagchi2013,Li2019}.
We aim to investigate how non-Hermitian interactions affect spin dynamics in the semiclassical Hubbard model at $T \to \infty $.

In our problem, there are $ 6N $ equations of motion for the spin components [Eq.~\eqref{classical_eoms}], but only $4N$ independent variables due to the conservation of spin norm at each site. 
We found that non-Hermitian interactions drive the system toward the $ZZ$ steady state, and simulations with periodic spin ladders of length $\mathcal{O}(10^2)$ are sufficient to capture this evolution.
The $2N$ spins are initialized randomly by uniformly sampling $2N$ points on the unit Bloch sphere. 
The $6N$ equations are then solved numerically using the Rosenbrock-Wanner Rodas5P method, implemented in the \texttt{DifferentialEquations.jl} package, with a time step of $0.01-0.05$. 
We verified that the total spin norm, i.e., $ \sum_j |\bm{s}_j|/N $, is well conserved during the time evolution, with a precision of $ 10^{-9}$. 
Averages over spin components or correlation functions are usually taken over an ensemble of $ 10^4 - 10^5 $ initial conditions.

\subsection{Steady state rung magnetization}

Although the $ T \to \infty $ problem is typically explored using numerical methods, some conclusions can be drawn by extrapolating from the uniform problem (Sec.~\ref{sec:uniform}). 
Given the randomness of the initial conditions, the $XY$ SS, which relies on a specific balance of $z$ components [$s^z(0) = \pm \tau^z(0)$], will be unstable.
An inspection of Eqs.~\eqref{classical_eoms} reveals that $ d\im\mathcal{H}/d \bm{s}_j \propto \tau^z_j \hat{z} $. 
Therefore, the spin-torque term induced by non-Hermitian interactions tends to polarize the spins along the $z$ axis. Even with small imaginary interactions, the damping term becomes dominant over time, leading the system to flow towards the dimerized $ZZ$ SS, either in the ferromagnetic or antiferromagnetic configuration, depending on the sign of the imaginary interactions.
When the imaginary interactions are large, $U|\sin\phi| \gg J$, the similarity to the uniform case becomes even more pronounced.
Neglecting the weaker hopping terms $J$ effectively decouples the rungs. The evolution of $ s^z_j $ and $ \tau^z_j $ on each rung follows the same equations as in the uniform case, resulting in similar $ZZ$ steady states.

\begin{figure}[t]
\includegraphics[width=\columnwidth]{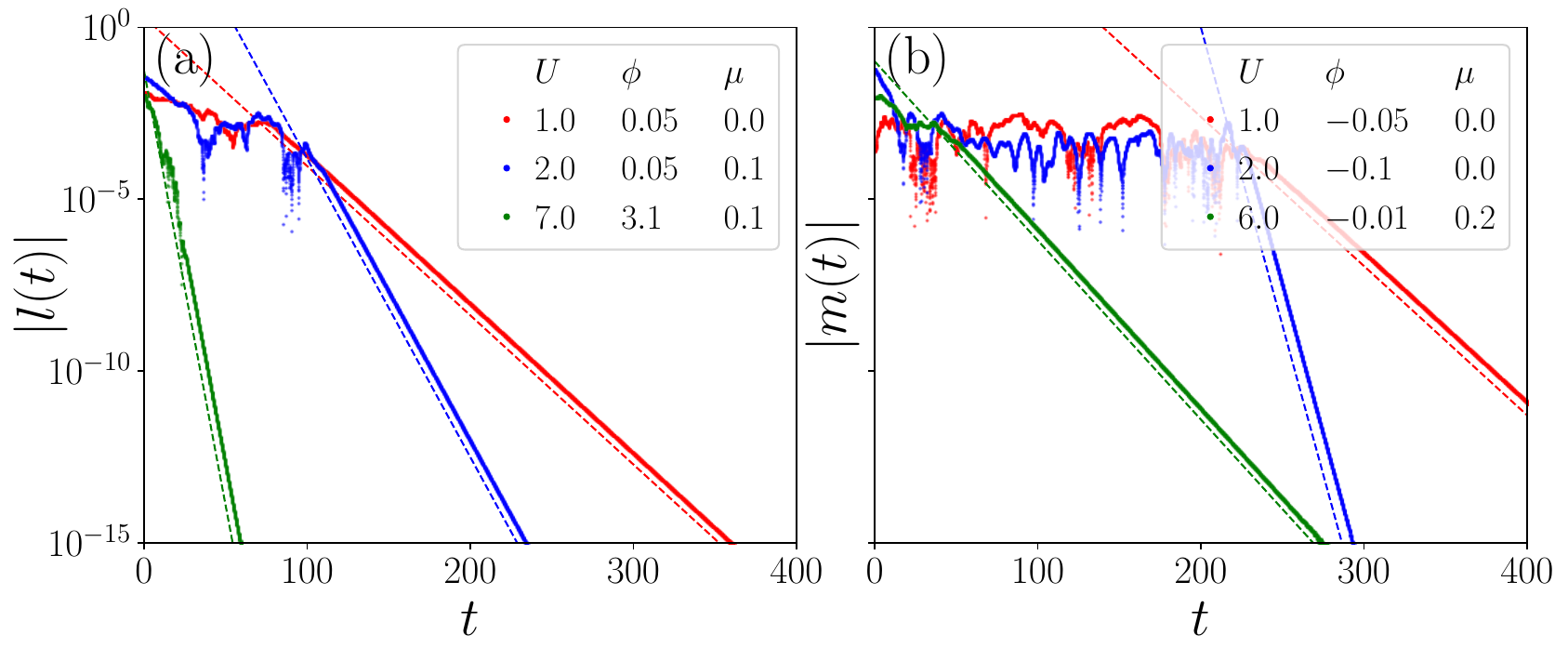}
\caption{$(a)$ The total staggered magnetization $l(t)$ at $\sin\phi>0$ and $(b)$ the total magnetization $m(t)$ at $\sin\phi<0$ obey the decay law of Eq.~\eqref{equator_ml} (dashed line) to the ferromagnetic point and to the antiferromagnetic point, respectively.
The simulations are performed for $N=100$ length spin ladder with random initial conditions, and three typical parameter sets in each case.
}
\label{fig:stag_mag}
\end{figure}

Specifically, as in the uniform case, the system evolves towards the ferromagnetic $ZZ$ SS for $ \sin\phi > 0 $, with the total staggered magnetization $ l(t) $ decaying exponentially as described by Eq.~\eqref{equator_ml}. 
For $\sin\phi < 0$, the system flows towards the antiferromagnetic $ZZ$ SS, with the total magnetization $m(t)$ exhibiting the same exponential decay. 
These exponential behaviors are confirmed by numerical simulations, as demonstrated in Fig.~\ref{fig:stag_mag} for various initial conditions.

Given that the steady state involves spins with only non-vanishing $s^z$ and $\tau^z$ components, we can characterize the system's approach to the steady state by defining and tracking the total polarization along the $z$ axis,
\begin{equation}
    p_s(t) = \frac{1}{N}\sum_j |s^z_j(t)|,
\end{equation}
and similarly for $p_\tau(t)$ for the $\tau$ spins. 
Since both spins follow identical dynamics, $p_s$ behaves the same as $p_\tau$, so it suffices to focus on the $s$ spins.

\begin{figure}[th]
\includegraphics[width=\columnwidth]{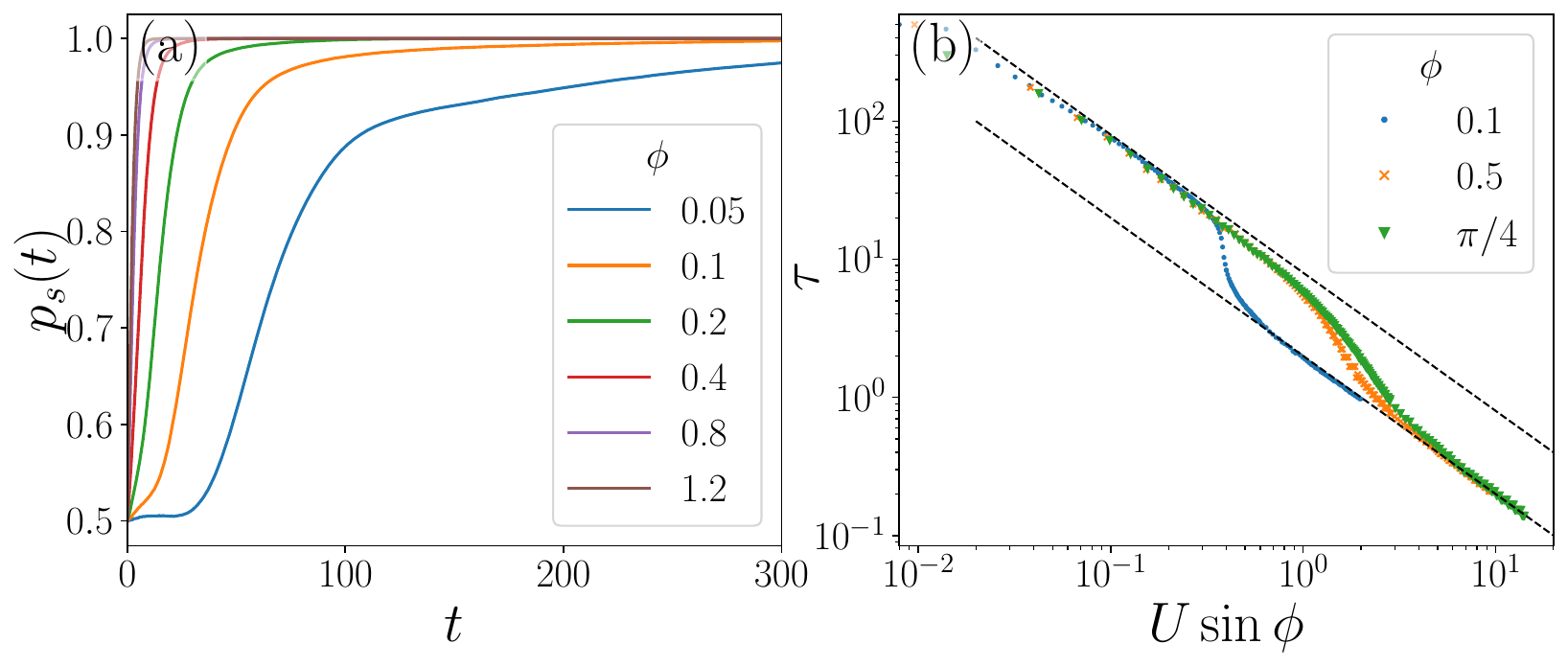}
\caption{(a) Typical total polarization $p_s(t)$ for different $\phi$ at $U=\pi/4$, showing the approach to steady state $p_s(t_\infty)=1$ 
[$N=200$ and $10^4$ random initial conditions].
Standard error bars are not visible.
(b) The characteristic timescale $\tau$ for approaching the steady state for several $U$ and three $\phi$ values decays as $1/U\sin\phi$ (dashed black lines) [$N=200$ and $10^3$ random initial conditions].
}
\label{fig:polariz}
\end{figure}

A typical behavior is shown in Fig.~\ref{fig:polariz}(a), where the total polarization reaches the steady-state value $p_s(t_\infty)=1$. 
The steady state is approached exponentially after some time $t_0$,
\begin{equation}\label{tau}
    1-p_s(t)=[1-p_s(t_0)]e^{(t_0-t)/\tau},
\end{equation}
where $\tau$ is the characteristic timescale to reach the steady state. The time $t_0$ can be chosen to minimize the effects of an initial transient regime, during which spins oscillate under weak interactions.

To determine $\tau$, we calculated $p_s(t)$, averaged over $10^3$ initial conditions, for several $U$ values and three different $\phi$ values. As expected, $\tau$ is proportional to $1/U\sin\phi$, consistent with the decay laws in Eqs.~\eqref{equator_ml} [see Fig.~\ref{fig:polariz}(b)]. 
However, $\tau$ exhibits a jump between two such $1/\sin\phi$ decays, which is related to a phenomenon discussed in the next subsection, the formation of ferromagnetic domains along the spin ladder. 
This jump in $\tau$ occurs when the domain size shrinks to the size of the lattice constant, at which point the spins become completely disordered in the longitudinal direction along the chain.

Finally, note that observing the steady states in the semiclassical problem is also conditioned by an additional timescale $\tau'$ stemming from the quantum problem, in the specific case where the Hamiltonian~\eqref{FH_spin} follows from the quantum master equation within the quantum trajectory approach [App.~\ref{sec:lindblad}].
In the latter theory, the effective non-Hermitian model governs the system evolution up to the timescale $\tau'$ associated with the occurrence of quantum jumps.
Thus, to ensure that the semiclassical system reaches the steady state, it is necessary that $\tau\lesssim\tau'$. This condition can always be satisfied for sufficiently large imaginary interactions $U\sin\phi$, corresponding to strong two-body dissipation rates.

\subsection{Longitudinal ferromagnetic domains}
The analysis of the uniform case does not fully capture the behavior in the infinite-temperature limit. 
While it helps us conclude that the steady states are fully dimerized, with spins within each rung locked in either a ferromagnetic or antiferromagnetic configuration, it does not provide information about possible spin ordering along the longitudinal direction of each chain.

In the strongly interacting regime, $U|\sin\phi|\gg J$, each rung is effectively decoupled from its neighbors, and as it reaches its steady state, it remains largely uncorrelated with adjacent rungs. To illustrate, consider the case where $\sin\phi>0$. Here, the steady state is characterized by $l_j(t_\infty)=0$, while $m_j(t_\infty)$ takes values of either $1$ or $-1$, depending on the initial conditions of each rung. 
Since these initial conditions are random, $m_j$, and similarly $s^z_j$ and $\tau_j$, flip randomly between $\pm 1$ along the chain, leading the total magnetization averaged over all sites to approach zero.

\begin{figure}[th]
    \includegraphics[width=\columnwidth]{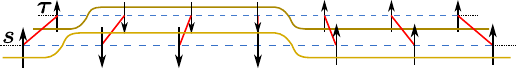}
    \includegraphics[width=\columnwidth]{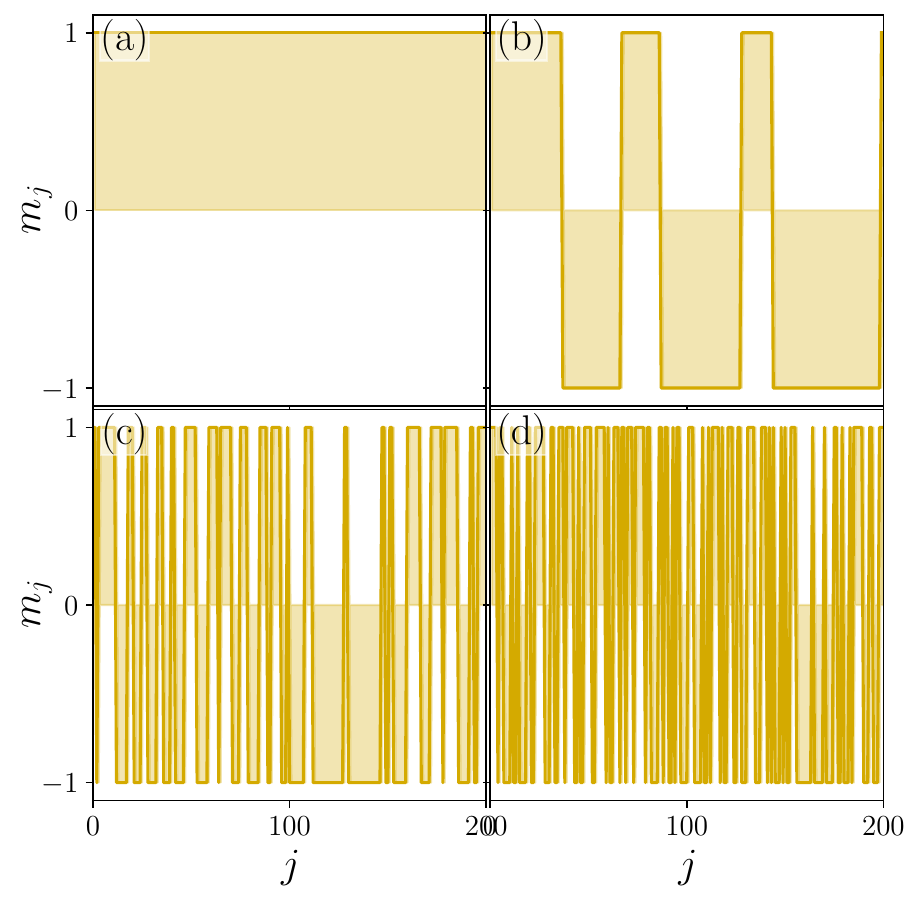}
    \caption{(Top) Sketch of domain-walls (dark yellow lines) in the steady state of  the $T\to\infty$ model. Here at $\sin\phi>0$, the spins are ferromagnetically coupled in each rung along the $z$ axis. 
    Similar phenomenology is seen at $\sin\phi<0$, where the rungs are antiferromagnetically coupled.
    (Bottom) 
    Measured rung magnetization $m_j$ in the steady state reveals domain walls at (a) $U=0.1$, (b) $U=1$, (c) $U=3$, and (d) $U=6$. Results are obtained for a single run at $T\to\infty$, $N=200$, and $\phi=\pi/4$.}
    \label{fig:domain}
\end{figure}

In contrast, the hopping terms $J$ play a role during the transient regime, leading to synchronization of spins in both the $\bm s$ and $\bm\tau$ chains. 
This synchronization results in the formation of ferromagnetic domains along each chain, as illustrated in Fig.~\ref{fig:domain}(top), for both dimerized phases. 
Therefore, regardless of the transverse or rung coupling between the spins, the steady state in the longitudinal direction will consist of domains of spins polarized either up or down along the $z$ axis.

For weak enough interactions, the average domain size can become larger than the system size in simulations, leading to the formation of a single domain spanning the entire system [see Fig.~\ref{fig:domain}(a)]. 
As the strength of the imaginary interactions increases, the spin domains along the chain gradually shrink until they reach the size of the lattice constant, effectively turning the system into uncorrelated dimers.
Snapshots of this evolution are shown in Fig.~\ref{fig:domain}(a-d) for different interaction strengths $U$ and with $\phi=\pi/4$. 
Given that $\sin\phi>0$, the steady state is characterized by ferromagnetic rungs, and we measure the rung magnetization $m_j$, which corresponds to both $s^z_j$ and $\tau^z_j$. Domain walls are identified by abrupt jumps in the magnetization between 1 and $-1$. 
A similar behavior is observed for $\sin\phi<0$, where the rungs are antiferromagnetically ordered, leading to the formation of analogous domains in the staggered magnetization $l_j$.

\begin{figure}[th]
    \includegraphics[width=\columnwidth]{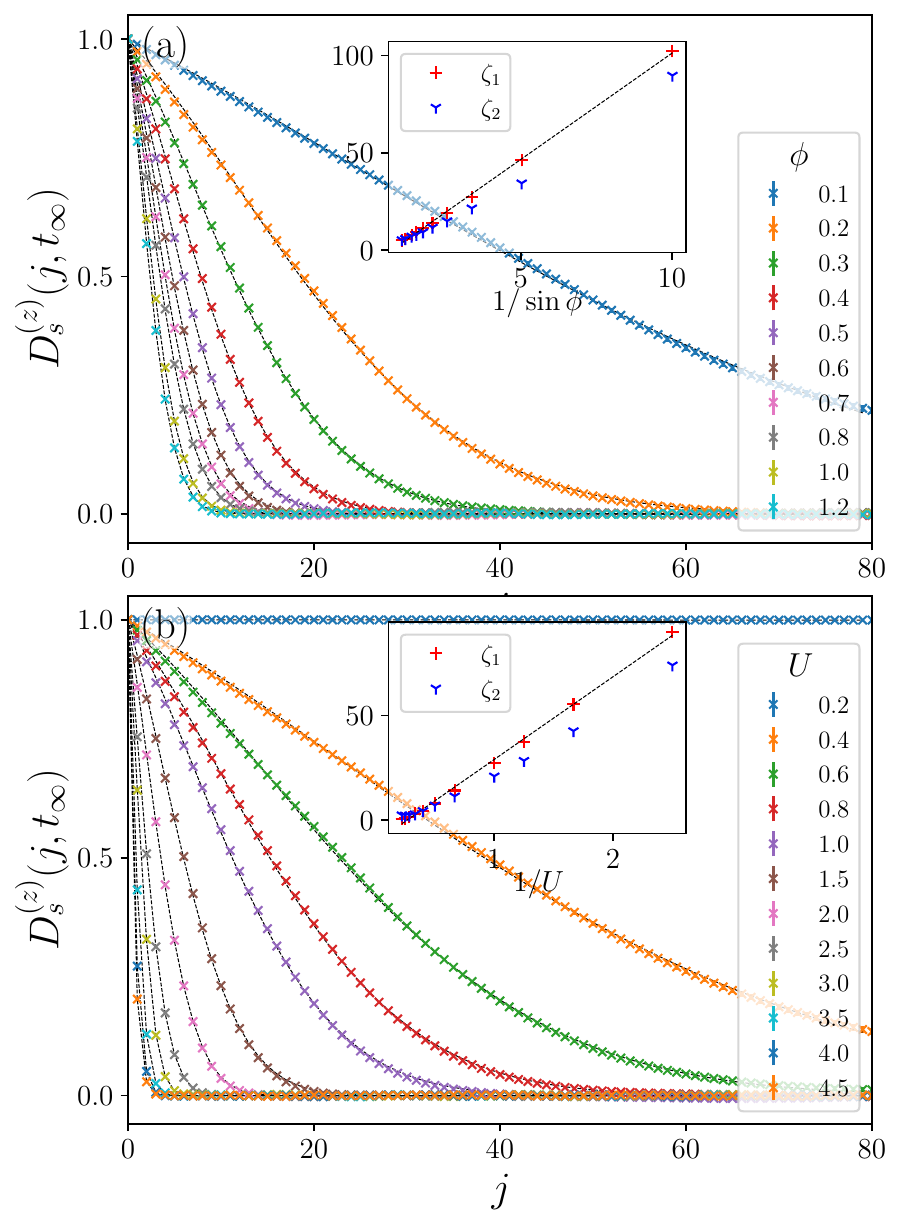}
    \caption{Equal-time correlation function~\eqref{D_corr} in the steady state $t_\infty$ as a function of distance $j$, fitted with Eq.~\eqref{D_inf} (dashed black line), for (a) different phases $\phi$, and $U=2$, and (b) $\phi=\pi/4$, but different interaction strengths $U$.
    The insets show the decay lengths $\zeta_1$ and $\zeta_2$ as a function of (a) $1/\sin\phi$ and (b) $1/U$, and a linear fit for $\zeta_1$. Standard errors are plotted, but are not visible.
    [$N=200$, and $10^4$ random initial conditions.]}
    \label{fig:eqCorr}
\end{figure}

For all interactions where $ U\sin\phi > 0 $, the total magnetization $m$ tends to zero when averaged over a large ensemble of initial conditions. 
Even if a single domain forms at weak interactions, approximately half of the samples will have $ m = 1 $ and the other half $ m = -1 $. 
Since $ l = 0 $ due to ferromagnetic rungs, both magnetizations $ m $ and $l$ are zero in the steady state. Similar arguments apply when $ U\sin\phi < 0 $: $ m = 0 $ due to antiferromagnetic rungs, and $ l = 0 $ due to ensemble averaging. However, domain formation when starting from an ensemble of initial conditions is revealed by the equal-time correlation functions,
\begin{equation}\label{D_corr}
D^{(z)}_s(j,t) = \frac{1}{N}\sum_i\avg{s^z_i(t)s^z_{i+j}(t)}.
\end{equation}
Similarly, $D^{(z)}_\tau(j,t)$ is defined for the $\bm\tau$ spins. Since the equations of motion are symmetric under the interchange of spin type, the two correlation functions, averaged over a large ensemble of initial conditions, carry the same information. Therefore, we focus on $D^{(z)}_s$ in the following.

The equal-time correlation function $\eqref{D_corr}$ in the steady state captures the decay of correlations along each chain, providing insight into the size of ferromagnetic domains in the system. Two examples are shown in Figs.~\ref{fig:eqCorr}(a) and~\ref{fig:eqCorr}(b), where either the phase $\phi$ or the interaction strength $U$ is varied. Due to periodic boundary conditions, the correlation functions for $j \geq N/2$ are identical to those for $j < N/2$. The correlations are maximal and constant when a single domain forms in the entire system at weak imaginary interactions [e.g., Fig.~\ref{fig:eqCorr}(b) for $U=0.2$, $\phi = \pi/4$, and $N=200$]. 
Generally, as expected, correlations~\eqref{D_corr} decay for $j < N/2$ as $\sin\phi$ or $U$ increases.

Notably, the decay cannot be fitted with a single exponential law due to the accelerated decay observed at large distances. 
Instead, the data fits well with 
\begin{equation}\label{D_inf}
D^{(z)}_s(j,t_\infty) = \exp\left(-\frac{j}{\zeta_1} - \frac{j^2}{\zeta_2^2}\right),
\end{equation}
which includes $j^2$ corrections. The insets of Fig.~\ref{fig:eqCorr} show the characteristic lengths $\zeta_1$ and $\zeta_2$, which prove to be of the same order of magnitude. In this case, the size of the ferromagnetic domain can be approximated by $\zeta_1$ alone. Moreover, $\zeta_1$ exhibits an approximately linear decay with the imaginary interaction strength $U\sin\phi$. The decay law~\eqref{D_inf} is inexact for very large $U\sin\phi$, where the domain size is on the order of the lattice constant, and the data to fit consists of just distances $j=0$ and $1$.

In summary, ferromagnetism along the chains arises from $XY$ interactions in the longitudinal direction, while ferro- and antiferromagnetism within a dimer are driven by the polarizing effect of the damping term. Note that the straightforward translation of these classical results to the quantum case ($S=1/2$) is not warranted as quantum effects were neglected in the classical approach. For example, such construction does not explain the formation of highly-entangled Dicke states in the quantum case~\cite{FossFeig2012, Sponselee2018}.


\subsection{Dynamical spin correlation functions}\label{sec:dynamic}
\begin{figure}[t]
    \includegraphics[width=\columnwidth]{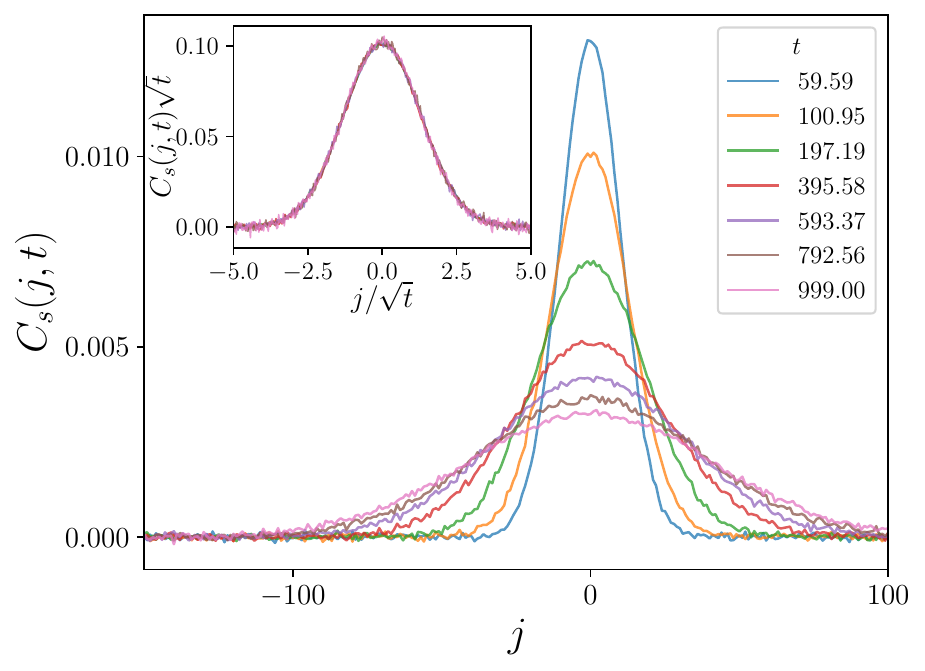}
    \caption{Spin correlation functions in the Hermitian limit $\phi=0$ obey a diffusive scaling law, collapsing on a unique universal Gaussian function when scaled with $\sqrt t$ (inset). Each chain has $N=10^3$ spins, $U=4$, and results are averaged over $10^5$ random initial conditions.
    }
    \label{fig:corr_diffusive}
\end{figure}
Spin dynamics in both classical and quantum systems is typically analyzed by examining the evolution of spin autocorrelation functions. In the infinite temperature limit, the dynamics of a classical Heisenberg spin chain is diffusive, and the system never reaches a steady state. Instead, the autocorrelation decays following a power law in the long-time limit, $\avg{\bm{S}_i(t) \cdot \bm{S}_i(0)} \sim t^{-1/2}$~\cite{Gerling1990,Bagchi2013}. 
In contrast, non-Hermitian systems, like the one considered here, evolve towards a steady state, causing the correlation functions to eventually saturate, and become time-independent. 
However, we demonstrate that the dynamics driven by the spin-transfer torque term exhibits anomalous behavior. 
We focus on the following dynamic spin correlation functions:
\begin{equation}\label{correlations}
C_s(j,t) = \frac{1}{N}\sum_i\avg{\bm s_i(0)\cdot \bm s_{i+j}(t)},
\end{equation}
and a similar one $C_\tau(j,t)$, defined  for the $\bm\tau$ spins. 
In our numerical calculations, we usually perform the average over $10^4-10^5$ random initial conditions. 
Due to the symmetry under exchange of $\bm s_j$ and $\bm\tau_j$ spins in Eqs.~\eqref{classical_eoms}, the correlations functions $C_s$ and $C_\tau$ behave identically, and we show only the results for $C_s$ in the following.

\begin{figure}[t]
    \includegraphics[width=\columnwidth]{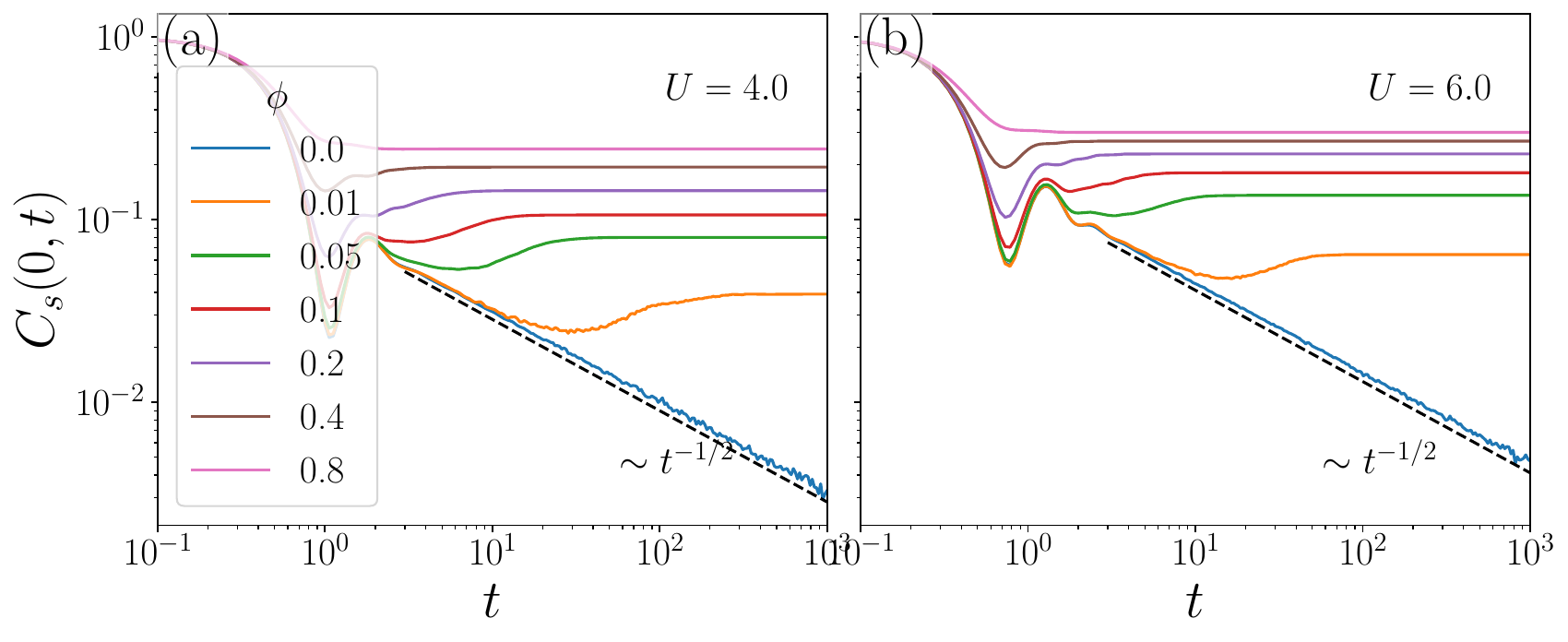} 
    \caption{Autocorrelation functions $C_s(0,t)$ for different non-Hermitian phases $\phi>0$ and $N=200$ at (a) $U=4$ and (b) $U=6$.
    At $\phi=0$, $N=1000$, the diffusive propagation of correlations follows the expected $t^{-1/2}$ law (dashed black line).
    Averages are over $10^4$ random initial conditions at each $\phi$.}
    \label{fig:timeCorr}
\end{figure}

In the absence of imaginary interactions, the spin dynamics is given by the Landau-Lifshitz equations~\eqref{classical_eoms} with $\im \mc H=0$.
We have scanned for several interaction strengths $U$ and found that the correlations indicate diffusive evolution of spins, similar to the classical Heisenberg chain.
In the long-time limit, the hydrodynamic regime is reached, with correlation functions showing diffusive scaling,
\begin{equation}\label{scaling}
C_s(j,t) \sim t^{-1/2}f({jt^{-1/2}}),
\end{equation}
with a universal Gaussian function $f$.
The correlation functions at several times $t$ are also displayed in Fig.~\ref{fig:corr_diffusive}, showing an excellent collapse on the Gaussian function $f$ according to Eq.~\eqref{scaling}.
The same result is also supported by studying the autocorrelation function $C_s(0,t)$, which exhibits the expected diffusive decay with $t^{-1/2}$ in Fig.~\ref{fig:timeCorr}, for two interactions strengths.

Let us now switch on the imaginary interactions. The system evolves towards an $ZZ$ steady state due to the non-Hermitian spin-torque term. The larger the imaginary interactions, the faster the system reaches this steady state. In our simulations, we set the amplitude of $U\sin\phi$ large enough so that convergence to the steady state is achieved for chains with $N \sim \mc{O}(10^2)$ spins and for times $t < 10^4$. Typical results for the spin autocorrelation functions $C_s(0,t)$ are shown in Fig.~\ref{fig:timeCorr} for two interaction strengths $U$ and various phases $\phi > 0$. Unlike the Hermitian case at $\phi = 0$, the correlation functions become constant in time once the steady state is reached. Additionally, for weak imaginary interactions (e.g., $\phi = 0.01$ in Fig.~\ref{fig:timeCorr}), the short-time behavior of correlations initially follows the typical diffusive scaling $t^{-1/2}$ before the non-Hermitian effects take over, eventually driving the system to the  steady state.

In Fig.~\ref{fig:timeCorr} it is also clear that the correlations in the steady state $C(0,t_\infty)$ generally increase with the $U\sin\phi$. 
This is easy to understand, although harder to quantify. For stronger interactions, the spins reach the steady state fast, and their final orientation is influenced by their initial random orientation. 
For example, if a pair with $\bm s_j$ and $\bm \tau_j$ pointing somewhere in the northern part of the Bloch sphere, then for $\sin\phi>0$ it will tend to evolve to a ferromagnetic alignment with $s^z_j=\tau^z_j=1$, so the final state is more correlated to the initial state. At smaller $U\sin\phi$ this effect is less pronounced as interactions along the chain might flip the orientation of these spins to $s^z_j=\tau^z_j=-1$.

\begin{figure}[t]
    \includegraphics[width=\columnwidth]{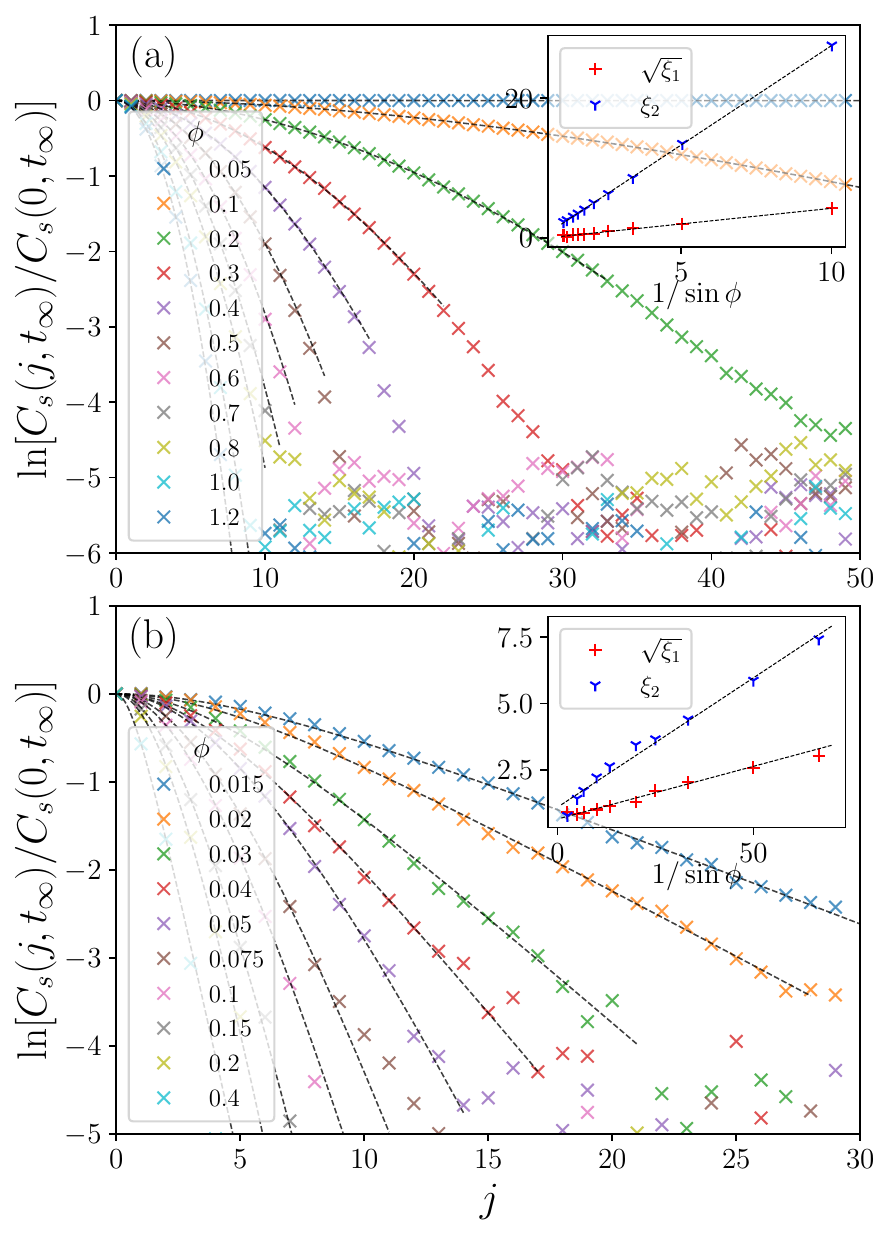}
    \caption{Spatial dependence of the spin correlation functions in the steady state of the non-Hermitian model at (a) $U=2$ and (b) $U=4$ for several phases $\phi$. 
    The correlation decay is no longer Gaussian, and it is fitted with the two-parameter Gumbel function~\eqref{gumbel}. 
    The parameters controlling the decay $\sqrt{\xi_1}$ and $\xi_2$ are plotted in the inset as a function of $1/\sin\phi$.}
    \label{fig:gumbel}
\end{figure}

These frozen correlations carry the signature of the anomalous spin dynamics in the transient regime, before reaching the steady state.
In Fig.~\ref{fig:gumbel}, we show the normalized correlations in the steady state.
These no longer follow the diffusive Gaussian behavior of the Hermitian case (Fig.~\ref{fig:corr_diffusive}), but they show an exponential decay at large distance.
This observation leads us to propose an ansatz for the frozen correlations in the form of the Gumbel distribution function,
\begin{equation}\label{gumbel}
\ln\left(\frac{C_s(j,t_\infty)}{C_s(0,t_\infty)}\right) 
= -\frac{j}{\xi_1}+\frac{\xi_2^2}{\xi_1^2}(1-e^{-j\xi_1/\xi_2^2}).
\end{equation}
The Gumbel $C_s(j,t_\infty)$ function interpolates between the Gaussian behavior $C_s(j,t_\infty)\simeq \exp(-\frac12j^2/\xi_2^2)$ seen at short distances, $j\ll \xi_2^2/\xi_1$, and the exponential decay $C_s(j,t_\infty)\simeq \exp(-j/\xi_1)$, at large distances, $j\gg \xi_2^2/\xi_1$.

In Fig.~\ref{fig:gumbel} it is shown that as the imaginary interaction strength grows, the spread of correlations is arrested sooner since the steady state is reached faster. 
For very small phases $\phi$, the correlations had time to decay and spread almost equally over the window of sites shown, leading to almost flat correlations, but very small in amplitude [see $\phi=0.05$ in Fig.~\ref{fig:gumbel}(a), with the normalization with $C_s(0,t_\infty)$ hiding the weak amplitude].
At larger interactions, the correlations have time to spread only over a few sites before being frozen in the steady state. 
In such cases, one might suspect that the hydrodynamic regime is not reached before the system reaches the steady state.
From a two-parameter fit, the decay lengths $\xi_1$ and $\xi_2$ as a function of the imaginary interaction strength are extracted and displayed in the insets of Fig.~\ref{fig:gumbel}.
The almost linear dependence of $\sqrt{\xi_1}$ and $\xi_2$ on $1/\sin\phi$
suggests that in both short and large-distance limits, the correlations decay as 
\begin{equation}\label{sin2phi}
\ln\left(\frac{C_s(j,t_\infty)}{C_s(0,t_\infty)}\right) 
\simeq [-j+c(1-e^{-j/c})]\sin^2\phi,
\end{equation}
with $c$ some constant that depends weakly on $\phi$.
However, at large $U\sin\phi$, $c$ show random fluctuations since the correlations had only a short time evolution and a spread of only a few sites before freezing in the steady state.
The dependence of correlations on $\sin^2\phi$ instead of $\sin\phi$ in Eq.~\eqref{sin2phi} is a surprise and needs further investigation.

\section{Discussion and Conclusion}
\label{sec:conclusion}
We have investigated spin dynamics in a non-Hermitian classical spin ladder model constructed as the semiclassical limit of the Fermi-Hubbard model with imaginary interactions. 
This construction neglects quantum effects~\cite{Nowak2007}, and the results obtained from the semiclassical model might not directly translate to the quantum model.
Nonetheless, the semiclassical model is intriguing in its own right. We observed that non-Hermitian interactions exert a damping effect on the spin equations of motion, similar to the Slonczewski spin-transfer torque. 
In the long-time limit, this damping term dominates, ensuring the system reaches a steady state. When spins are initialized in a random but uniform state, a steady state of decoupled spin chains can emerge. However, for general initial conditions in the infinite-temperature limit, the system favors dimerized steady states where all rungs in the ladder are decoupled. 
Depending on the sign of the imaginary interactions, spins in a rung become locked in either ferromagnetic or antiferromagnetic order along the $z$ axis. 
Additionally, the steady state shows the formation of ferromagnetic domains along each chain. 
These domains grow under weak imaginary inter-chain interactions, suggesting they result from intra-chain coupling relevant during the transient regime.

Dynamic correlation functions reveal signatures of anomalous dynamics in the non-Hermitian system. In the Hermitian limit, the ladder model displays normal diffusive spin dynamics. Introducing the damping term allows the system to achieve a steady state where the correlation functions are ``frozen.'' These functions deviate from the typical Gaussian profile associated with diffusive dynamics, instead showing exponential decay at larger distances.

Applying these ideas to other non-Hermitian quantum spin systems~\cite{Shibata2019,Yamamoto2022} and studying the crossover from the fully quantum to the classical limit in a controlled manner would be interesting. Additionally, introducing topology to the single-particle band structure and examining its effect in the classical limit holds promise. Lastly, considering the first quantum corrections and the impact of magnons on classical solutions and dynamics could provide crucial insights into the stability of the classical limit and the role of quantum fluctuations~\cite{Tserkovnyak2020,Flebus2020,Hurst2022,Zou2024,Yu2024}.

\begin{acknowledgments}
This work received financial support from the Romanian MCID through the ``Nucleu'' Program within the PNCDI IV 2022-2027, project PN 23 24 01 04.
This research was also supported by the National Research, Development and
Innovation Office - NKFIH  within the Quantum Technology National Excellence
Program (Project No.~2017-1.2.1-NKP-2017-00001), K134437, K142179, by the BME-Nanotechnology
FIKP grant (BME FIKP-NAT).
\end{acknowledgments}

\appendix
\onecolumngrid

\section{Derivation of the non-Hermitian Hubbard Hamiltonian}
\label{sec:lindblad}
This appendix reviews the derivation of the quantum non-Hermitian Hubbard Hamiltonian~\eqref{FH} from the quantum master equation with two-body losses.

Consider a fermionic system described by the Hermitian Hubbard Hamiltonian
\begin{equation}
H_0=-\mc J \sum_{j\sigma} (c^\dag_{j\sigma} c^{}_{j+1\sigma}+\hc)
-\text{\textmu}\sum_j(n_{j\up}+n_{j\down})+\mc U'\sum_j (n_{j\up}-\frac12) (n_{j\down}-\frac12),
\end{equation}
which is coupled to an external reservoir allowing for one-body gain and two-body losses.
The evolution of the system's density matrix is governed by the Lindblad equation
\begin{equation}
\dot\rho(t) = -i[H_0,\rho(t)] + \sum_{\a=\{\up,\down,\up\down\}}^{}\sum_{j=1}^{N} L_{\a,j}\rho(t) L_{\a,j}^\dag -\frac12\{L_{\a,j}^\dag L_{\a,j},\rho(t)\}. 
\end{equation}
The Lindblad jump operators act at each site and reflect the single-particle gains, and two-particle losses are
\begin{equation}
L_{\up,j}=\sqrt{\gamma}c_{j\up}^\dag,
\quad
L_{\down,j}=\sqrt{\gamma}c_{j\down}^\dag,
\quad L_{\up\down,j} = \sqrt{2\gamma} c_{j\up}c_{j\down}.
\end{equation}
The rates $\gamma$ for the one-body gains are half the rates for two-body losses. 
Following the quantum trajectory method~\cite{Daley2014}, the evolution of the system in the absence of quantum jumps is governed by the effective non-Hermitian Hamiltonian
\begin{align}
H &= H_0 - \frac{i}{2}\sum_{\a=\{\up,\down,\up\down\}}^{}\sum_{j=1}^N L^\dag_{\a,j} L_{\a,j} = H_0 - i\gamma \sum_{j=1}^N (n_{j\up}-\frac12) (n_{j\down}-\frac12),
\end{align}
where the Hamiltonian was shifted by a constant. Further on, denoting $\mc U=\sqrt{{\mc U'}^2+\gamma^2}$ and $\tan\phi = -\gamma/\mc U'$ yields the starting Hamiltonian in Eq.~\eqref{FH} in the main text. The repulsive Hubbard model is described by $\phi<0$, and the attractive one by $\phi>0$.
In ultracold atoms experiments the phase $\phi$ can be tuned by varying $\mc U'$, controlling the scattering length for elastic collisions by Feshbach resonances~\cite{Chin2010}. Alternatively, it is possible to directly control the two-body loss rate through photoassociation, as was shown in Ref.~\cite{Tomita2017} in a 1D quantum gas of \ce{^{174}Yb} atoms.

Note that the presence of one-body gains and the tuning of their rate is not crucial. 
It was only considered since it yields simpler quantum spin equations. In the absence of one-body gains, the spin model~\eqref{FH_spin} would gain linear terms in spins, which can be further absorbed in a complex chemical potential.

\section{Spin coherent basis}
\label{sec:coherent}
We consider the following variational ansatz for the semiclassical problem~\cite{Schliemann1998}. The total wave function of the system is a direct product of coherent states at each site,
\begin{equation}
|\psi(t)\rangle = \bigotimes_{j=1}^N (|S_j;\theta_j(t),\varphi_j(t)\rangle\otimes|T_j;\theta'_j(t),\varphi'_j(t)\rangle),
\end{equation}
The coherent states at an arbitrary site (with index removed for improved legibility) are
\begin{equation}\label{basis}
|S;\theta,\varphi\rangle = e^{-i\varphi S^z}e^{-i\theta S^y}|S\rangle,
\end{equation}
with a similar expression for $T$ spins. Single-spin expectation values in the coherent state~\eqref{basis} follows readily,
\begin{equation}
\avg{\bm n\cdot \bm S} = S, \quad \bm n = (\sin\theta\cos\varphi,\sin\theta\sin\varphi,\cos\theta). 
\end{equation}
The expectation values behave as classical spins of length $S$. 
Semiclassical corrections occur in two-spin expectation values, where non-commutativity of spins at the same site yields
\begin{eqnarray}\label{factApp}
\avg{S^\a_j S^\b_k} &=& \avg{S^\a_j}\avg{S^\b_k}(1-\frac{\delta_{jk}}{2S})+\delta_{\a\b}\delta_{jk}\frac{S}{2} +i\delta_{jk}\e_{\a\b\g}\frac{\avg{S^\g_j}}{2},
\end{eqnarray}
with $\gamma$ summed over $x,y,z$.

\section{Equations of motion}
The right-hand side of Eq.~\eqref{eom2} is calculated with the aid of Eqs.~\eqref{factApp}. 
The commutator and anticommutator yield,
\label{sec:eoms}
\begin{eqnarray}
\avg{[S^\alpha_j,H_+]} &=& -2iJ\e_{\alpha n\beta}\avg{S^\beta_j(S^n_{j+1}+S^n_{j-1})}
+i[U\cos(\phi) \avg{T_j^z}-\mu]\e_{\alpha z\beta}\avg{S^\beta_j},\notag\\
\avg{\{S^\alpha_j,\varGamma\}}-2\avg{S^\alpha_j}\avg{\varGamma} &=& 
2iU\sin(\phi) \bigg(
-\frac{\avg{S_j^\alpha}\avg{S^z_j}}{2S} + S \delta_{\alpha z}\bigg)\avg{T^z_j},
\end{eqnarray}
with index $\beta$ summed over $x,y,z$ and $n$ over $x,y$.
Back in Eq.~\eqref{eom2}, and using the unit classical spins $\bm s_j$ and $\bm\tau_j$~\eqref{unitspins}, yields the $6N$ semiclassical equations of motion explicitly shown here for each spin component,
\begin{eqnarray}\label{eom_phs}
    \dot s_j^x &=& -2Js^z_j(s_{j+1}^y+s_{j-1}^y) - U\cos(\phi)s_j^y\tau_j^z
    -U\sin(\phi) s_j^x s_j^z\tau_j^z+\mu s^y_j,
    \notag\\
    \dot s_j^y &=& 2Js_j^z(s_{j+1}^x+s_{j-1}^x)+U\cos(\phi) s_j^x\tau_j^z
    -U\sin(\phi) s_j^y s_j^z \tau_j^z-\mu s^x_j,
    \notag\\
    \dot s_j^z &=& -2J s_j^y(s_{j+1}^x+s_{j-1}^x) 
    + 2J s_j^x(s_{j+1}^y+s_{j-1}^y) - U\sin(\phi)(s_j^zs_j^z\tau_j^z - \tau_j^z),
\end{eqnarray}
and similarly for $\bm\tau_j$ spins under the interchange $s^\alpha_j\leftrightarrow\tau^\alpha_j$.
In vector notation, the system of equations takes the compact form presented in Eqs.~\eqref{classical_eoms}.

For uniform initial conditions, the equations reduce to 
\begin{align}\label{hom_6}
	\dot s^x &= -4 Js^ys^z- U\cos\phi s^y\tau^z- U\sin\phi s^xs^z\tau^z+\mu s^y,\\
	\dot s^y &= 4 Js^xs^z + U\cos\phi s^x\tau^z- U\sin\phi s^ys^z\tau^z-\mu s^x,\notag\\
	\dot s^z &= -  U\sin\phi[(s^z)^2-1]\tau^z,\notag
\end{align}
and similarly for $\bm\tau$ spins.

\twocolumngrid
\bibliography{bibl}
\end{document}